\documentclass[graybox]{svmult}

\usepackage{mathptmx}       
\usepackage{helvet}         
\usepackage{courier}        
\usepackage{type1cm}        
\usepackage{makeidx}         
\usepackage{graphicx}        
\usepackage{multicol}        
\usepackage[bottom]{footmisc}

\usepackage{amsmath,amssymb}
\usepackage{bm}             
\usepackage{bbm}            
\usepackage{mma1}

\def\rmd{\,\mathrm{d}}  
\def\D#1{D_{\!#1}}      
\def\S#1{S_{\!#1}}      
\def\d{\partial}
\def\Ann{\operatorname{Ann}}

\renewcommand{\leq}{\leqslant}

\def\F{\mathbbm F}
\def\K{\mathbbm K}
\def\K{\mathbbm K}

\def\OO{\mathbbm O} 
\def\N{\mathbbm N}
\def\Z{\mathbbm Z}
\def\Q{\mathbbm Q}

\def\0{{\bm 0}}
\def\1{{\bm 1}}

\def\k{{\bm k}}

\def\w{{\bm w}}
\def\x{{\bm x}}

\def\DD#1{{\bm D}_{\!#1}}
\def\SS#1{{\bm S}_{\!#1}}
\def\dd{\bm{\partial}}
\def\aalpha{\bm{\alpha}}

\def\pow{\mathbin{\raisebox{-2.5pt}{\hbox{\large$\,\hat{}\,$}}}}  

\renewcommand{\labelenumi}{(\arabic{enumi})}

\def\cmd#1{#1}
\def\Index#1{#1\index{#1}}

\makeindex             

\begin{document}

\title*{Creative Telescoping for Holonomic Functions}
\titlerunning{Creative Telescoping for Holonomic Functions}

\author{Christoph~Koutschan}
\institute{
C. Koutschan\\
Johann Radon Institute for Computational and Applied Mathematics (RICAM),
Austrian Academy of Sciences (\"OAW), Linz, Austria\\
\email{christoph.koutschan@ricam.oeaw.ac.at}
\bigskip\\
C. Schneider and J. Bl\"umlein (eds.), \emph{Computer Algebra in Quantum Field Theory},\\
Texts \& Monographs in Symbolic Computation, DOI 10.1007/978-3-7091-1616-6\_7,\\
Springer-Verlag Wien 2013
}

\maketitle

%
%

\abstract{
  The aim of this article is twofold: on the one hand it is intended
  to serve as a gentle introduction to the topic of creative telescoping, from
  a practical point of view; for this purpose its application to several
  problems is exemplified. On the other hand, this chapter has the flavour of
  a survey article: the developments in this area during the last two decades
  are sketched and a selection of references is compiled in order to highlight
  the impact of creative telescoping in numerous contexts.
}

\section{Introduction}
\label{sec.intro}

\index{holonomic|(}%
\index{creative telescoping!holonomic functions|(}%
The method of \emph{creative telescoping} is a widely used paradigm in
computer algebra, in order to treat symbolic sums\index{symbolic summation}
and integrals\index{symbolic integration} in an algorithmic way. Its modus
operandi is to derive, from an implicit description of the summand
resp. integrand, e.g., in terms of \Index{recurrence}s or \Index{differential
  equation}s, an implicit description for the sum resp. integral. The latter
can be used for proving an identity or for finding a closed form for the
expression in question. Algorithms that use this idea are nowadays implemented
in all major computer algebra systems. Meanwhile, they have been successfully
applied to many problems from various areas of mathematics and physics, see
Section~\ref{sec.appl} for a selection of such applications.

The key idea of creative telescoping is rather simple and works for summation
problems as well as for integrals. For example, consider the problem of
evaluating a sum of the form $F(n)=\sum_{k=a}^b f(k,n)$ for $a,b\in\Z$ and
some bivariate sequence~$f$. If one succeeds to find another bivariate
sequence~$g$ and univariate sequences~$c_0$ and~$c_1$ such that the equation
\begin{equation}\label{eq.ct1}
  c_1(n)f(k,n+1) + c_0(n)f(k,n) = g(k+1,n) - g(k,n)
\end{equation}
holds, then a \Index{recurrence} for the sum~$F$ is obtained by
summing~\eqref{eq.ct1}\index{summation!holonomic} with respect to~$k$
from~$a$ to~$b$, and then telescoping the right-hand side:
\[
  c_1(n)F(n+1) + c_0(n)F(n) = g(b+1,n) - g(a,n).
\]
For this reasoning to be nontrivial, one stipulates that the sequence~$g$ is
given as a closed-form expression in terms of the input (this will be made
precise later).  Note that on the left-hand side of~\eqref{eq.ct1} one can
have a longer linear combination of $f(k,n)$, \dots, $f(k,n+d)$, giving rise
to a higher-order recurrence for~$F$. This procedure works similarly for
integrals, see Section~\ref{sec.theory} for a detailed exposition. In order to
guarantee that a creative telescoping equation, like~\eqref{eq.ct1}, exists,
one requires that the summand~$f$ satisfies sufficiently many equations.  This
requirement leads to the concepts of \emph{\Index{holonomic functions}} and
\emph{$\d$-finite functions}\index{D-finite}\index{$\d$-finite}; they will be introduced in
Section~\ref{sec.dfinite}.

The class of \Index{holonomic functions} is quite rich and thus the method of
creative telescoping applies to a wide variety of summation and integration
problems.  Just to give the reader an impression of this diversity, we list a
random selection of identities that can be proven by the methods described in
this article (where $P_n^{(a,b)}(x)$ denotes the \Index{Jacobi polynomials}%
\index{orthogonal polynomials!Jacobi}, $L_n^a(x)$ the
\Index{Laguerre polynomials}\index{orthogonal polynomials!Laguerre}, $J_n(x)$ the Bessel
function\index{Bessel functions} of the first kind, $H_n(x)$ the
\Index{Hermite polynomials}\index{orthogonal polynomials!Hermite},
$C_n^{(\lambda)}(x)$ the \Index{Gegenbauer polynomials}%
\index{orthogonal polynomials!Gegenbauer}\index{ultraspherical polynomials},
$\Gamma(n)$ the \Index{Gamma function}, and $y_n(x)$ the spherical Bessel
function\index{spherical Bessel functions} of the second kind):

\begin{equation*}
  \sum_{k=0}^n\binom{n}{k}^2\binom{k+n}{k}^2=\sum_{k=0}^n\binom{n}{k}\binom{k+n}{k}\sum_{j=0}^k\binom{k}{j}^3,
\end{equation*}\index{binomial coefficient}
\begin{equation*}
  \int_0^\infty \frac{1}{\left(x^4+2 a x^2+1\right)^{m+1}}\rmd x=\frac{\pi
  P_m^{\left(m+\frac{1}{2},-m-\frac{1}{2}\right)}(a)}{2^{m+\frac {3}{2}}(a+1)^{m+\frac{1}{2}}},
\end{equation*}
\begin{equation*}
  \int_0^{\infty } e^{-t} t^{\frac{a}{2}+n} J_a\left(2 \sqrt{t x}\right)\rmd t = e^{-x} x^{a/2} n! L_n^a(x),
\end{equation*}
\begin{equation*}
  \sum _{n=0}^\infty\frac{(-t)^n y_{n-1}(z)}{n!} = \frac{1}{z}\sin \left(\sqrt{z^2+2 t z}\right),
\end{equation*}
\begin{equation*}
  \int_{-\infty}^\infty\sum_{m=0}^\infty\sum_{n=0}^\infty\frac{H_m(x)H_n(x)r^ms^ne^{-x^2}}{m!n!}\rmd x
  =\sqrt{\pi } e^{2 r s},
\end{equation*}
\begin{equation*}
  \int_{-1}^1\left(1-x^2\right)^{\nu-\frac{1}{2}} e^{i a x} C_n^{(\nu)}(x)\rmd x
  =\frac{\pi 2^{1-\nu} i^n \Gamma (n+2 \nu) a^{-\nu } J_{n+\nu }(a)}{n! \Gamma (\nu )}.
\end{equation*}
Further examples are discussed in Section~\ref{sec.demo} where we also
demonstrate the usage of our \Index{Mathematica} package HolonomicFunctions:%
\index{computer algebra system!Mathematica}%
\index{HolonomicFunctions package}\index{software package!HolonomicFunctions}
\begin{mma}
\In << |HolonomicFunctions.m| \\
\bigskip
\Print HolonomicFunctions package by Christoph Koutschan, RISC-Linz, \linebreak Version 1.6 (12.04.2012) \\
\end{mma}


For further reading, we recommend the following textbooks: the classic source
for \Index{hypergeometric} summation\index{summation!hypergeometric} is the wonderful book%
~\cite{PetkovsekWilfZeilberger96}, 
although \Index{Zeilberger's algorithm} made it already into the second edition of
Concrete Mathematics%
~\cite{GrahamKnuthPatashnik94}, 
as well as into its recent ``algorithmic supplement''%
~\cite{KauersPaule11}. 
A book that is completely dedicated to hypergeometric summation is%
~\cite{Koepf98}. 
We also would like to point the reader to the excellent survey articles%
~\cite{Cartier91, 
  Koornwinder93, 
  PauleStrehl95, 
  Zeilberger95, 
  Chyzak98b} 
and to the theses~\cite{Chyzak98,Koutschan09} for more detailed introductions
to the topic of creative telescoping in the context of \Index{holonomic functions}.

\section{History and Developments}
\label{sec.history}

The notion \emph{creative telescoping} was first coined by van der Poorten in
his essay~\cite{vanderPoorten79} on Ap\'ery's proof of the irrationality of
$\zeta(3)$\index{Apery sequence}\index{Riemann's zeta function}. But
certainly, the underlying principle was known and used long before as an ad
hoc trick to tackle sums and integrals. The most famous example is the
practice of \emph{differentiating under the integral sign}, that was made
popular by Feynman in his enjoyable book ``Surely You're Joking,
Mr. Feynman!''~\cite{Feynman85}, see also~\cite{AlmkvistZeilberger90}. It was
Zeilberger who equipped creative telescoping with a concrete well-defined
meaning and connected it to an algorithmic method%
~\cite{Zeilberger91}. 

The seminal paper that initiated all the developments presented here is
Zeilberger's 1990 \emph{holonomic systems approach}%
\index{Zeilberger's holonomic systems approach} paper\index{summation!holonomic}%
~\cite{Zeilberger90}. 
It sketches an algorithmic proof theory for identities among a large class of
elementary and \Index{special functions}, involving summation quantifiers and
integrals. The main theorems are based on the theory of
$D$-modules~\cite{Bernstein72,Coutinho95}, as well as the creative telescoping
algorithm which uses a general, but inefficient, elimination procedure.
Therefore, it was not really suited to be applied to real problems, except
from some toy examples, and was later called ``the slow algorithm''\index{Zeilberger's slow algorithm} by
Zeilberger, see Section~\ref{sec.slow}. But very quickly, one realized the big potential that lied in
these ideas. Takayama\index{Takayama's algorithm} designed a method that is
still based on elimination, but in a more sophisticated way using modules%
~\cite{Takayama90b}, 
see Section~\ref{sec.takayama}.
In the same year---we're still in 1990---more efficient creative telescoping
algorithms for special cases were formulated: Zeilberger's%
\index{Zeilberger's algorithm} celebrated ``fast algorithm'' for \Index{hypergeometric} single
sums\index{summation!hypergeometric}%
~\cite{Zeilberger90a} 
and its differential analogue, the \Index{Almkvist-Zeilberger algorithm} for the
integration of \Index{hyperexponential} functions\index{symbolic integration}%
~\cite{AlmkvistZeilberger90}. 
The theory on which these two algorithms are built was developed by Wilf and
Zeilberger%
~\cite{WilfZeilberger92} 
and was named \emph{WZ theory} after its inventors, who were awarded the Leroy
P. Steele Prize in 1998 for this seminal work.

In the following years the main focus of research in this field concentrated
on hypergeometric summation\index{summation!hypergeometric}. Certain extensions%
~\cite{Koepf95a} 
and optimizations%
~\cite{Riese01} 
of \Index{Zeilberger's algorithm} and its $q$-analogue\index{summation!$q$-hypergeometric}%
~\cite{PauleRiese97} 
were published. The problem of dealing with multiple sums\index{summation!multiple sums} was studied in more
detail%
~\cite{Wegschaider97, 
  ApagoduZeilberger06, 
  ChenHouMu06}, 
also for \Index{$q$-hypergeometric} terms\index{summation!$q$-hypergeometric}%
~\cite{Riese03}. 
Based on estimates on the order of the output \Index{recurrence} and the largest
integer root of its leading coefficient, Yen derived an a priori bound for the
number of instances one has to check in order to get a rigorous proof of a
($q$-) hypergeometric summation identity%
~\cite{Yen96, 
  Yen97}; 
although these bounds are too large for real applications, this in principle
allows to prove such identities by just verifying them on a finite set of
special cases, without executing \Index{Zeilberger's algorithm} explicitly.  This
bound was later improved drastically in%
~\cite{GuoHouSun08}. 
Sharp bounds for the order of the \Index{telescoper} that is computed by
Zeilberger's algorithm and its $q$-analogue were derived in%
~\cite{MohammedZeilberger05}. 
Abramov considered the question for which inputs the algorithm succeeds%
~\cite{AbramovLe02, 
  Abramov03}. 

In the late 1990s a return to the original ideas of Zeilberger started, namely
to consider general \Index{holonomic functions} instead of only ($q$-)
\Index{hypergeometric} / \Index{hyperexponential}
expressions\index{summation!holonomic}.  This development was initiated by
Chyzak and Salvy%
~\cite{ChyzakSalvy98, 
  Chyzak98} 
and culminated in a generalization of \Index{Zeilberger's algorithm} to holonomic
functions%
~\cite{Chyzak00} 
that is now known as \Index{Chyzak's algorithm}, see Section~\ref{sec.chyzak}.  This work was picked up
in~\cite{Koutschan09} where several nontrivial applications of creative
telescoping were presented. A fast but heuristic approach to the computation
of creative telescoping relations for general \Index{holonomic functions} was then
given in%
~\cite{Koutschan10c}, 
see Section~\ref{sec.heur}.

During the last few years, a new interest in creative telescoping algorithms
arose.  The main motivation was to understand the complexity of such
algorithms, a question that had been neglected during the two preceding
decades. This research finally also led to new algorithmic ideas. A first
attempt to study the complexity of creative telescoping was made in%
~\cite{BostanChenChyzakLi10}, 
but this investigation was restricted to bivariate rational functions as
inputs.  The problem of predicting the order and the degree of the
coefficients of the output was largely solved in%
~\cite{ChenKauers12a} 
for the \Index{hyperexponential} case and in%
~\cite{ChenKauers12b} 
for the hypergeometric case\index{summation!hypergeometric}. Both articles
also discuss the trading of order for degree, i.e., the option of computing an
equation with lower coefficient degree at the cost of a larger order and vice
versa; this trade-off can be used to reduce the complexity of the algorithms.
The question of existence criteria for creative telescoping relations for
mixed hypergeometric terms\index{hypergeometric!term} was answered in%
~\cite{ChenChyzakFengFuLi12}. 
Concerning new creative telescoping algorithms, the use of residues for the
computation of \Index{telescoper}s has been investigated in%
~\cite{ChenSinger12} 
for rational functions and in%
~\cite{ChenKauersSinger12} 
for \Index{algebraic functions}. Further innovations include an algorithm for
\Index{hyperexponential} functions based on Hermite reduction%
~\cite{BostanChenChyzakLiXin13} 
and new algorithm for rational functions%
~\cite{BostanLairezSalvy13} 
using the Griffiths-Dwork method.


Since our focus is on creative telescoping for holonomic functions, we mention
only briefly some other settings in which this method can be realized.  The
first algorithm for a class of non-holonomic sequences was given in%
~\cite{Majewicz96}, 
where Abel-type sums were considered.  An algorithm for summation of
expressions involving \Index{Stirling numbers} and similar non-holonomic bivariate
sequences was invented in%
~\cite{Kauers07}. 
Closure properties and creative telescoping for general non-holonomic
functions were presented in%
~\cite{ChyzakKauersSalvy09}. 
\index{creative telescoping!difference fields}
In the setting of difference fields, Schneider developed a sophisticated
\Index{symbolic summation} theory~\cite{Schneider01} whose core again is creative
telescoping.  For more information on this topic we refer to the book chapter%
~\cite{Schneider13}. 
Similarly, see~\cite{Raab12} for creative telescoping in differential fields.


We have already mentioned that algorithms based on creative telescoping are
part of many computer algebra systems. For example, Zeilberger's fast
algorithm%
~\cite{Zeilberger90a} 
for hypergeometric summation\index{summation!hypergeometric} has been implemented in
\Index{Maple}\index{computer algebra system!Maple}%
~\cite{Koornwinder93, 
  PetkovsekWilfZeilberger96}, 
shortly after its invention. In current Maple versions it is available by the
command SumTools[Hypergeometric][Zeilberger]. Other implementations of
\Index{Zeilberger's algorithm} are in \Index{Mathematica}%
\index{computer algebra system!Mathematica}%
\index{Fast Zeilberger package}%
\index{software package!Fast Zeilberger}
~\cite{PauleSchorn95}, 
in \Index{Reduce}\index{computer algebra system!Reduce}%
~\cite{Koepf95}, 
and in \Index{Macsyma}\index{computer algebra system!Macsyma}%
~\cite{Caruso00}. 
Its differential analogue, the \Index{Almkvist-Zeilberger algorithm}~\cite{AlmkvistZeilberger90},
can be called by DEtools[Zeilberger] in Maple.  For the $q$-analogue, 
Zeilberger's algorithm for \Index{$q$-hypergeometric}
summation\index{summation!$q$-hypergeometric}, there exist implementations in
Mathematica%
~\cite{Riese95,PauleRiese97} 
and in Maple%
~\cite{BoeingKoepf99}, 
see also the command QDifferenceEquations[Zeilberger] there.  Packages for
multiple sums\index{summation!multiple sums} have been written in Mathematica, namely MultiSum%
\index{software package!MultiSum}%
~\cite{Wegschaider97} 
for hypergeometric summands and its $q$-version qMultiSum%
\index{software package!qMultiSum}%
~\cite{Riese03} 
for \Index{$q$-hypergeometric} multi-sums. Multiple integrals\index{symbolic integration} can be treated with the
Maple package MultInt\index{software package!MultInt}%
~\cite{Tefera02}. 
Finally, there are two software packages for creative telescoping of general
\Index{holonomic functions}\index{summation!holonomic}, which are not restricted to
($q$-) hypergeometric / \Index{hyperexponential} inputs, i.e., expressions
satisfying first-order equations: Mgfun\index{software package!Mgfun}%
~\cite{Chyzak98} 
for \Index{Maple}\index{computer algebra system!Maple} and HolonomicFunctions%
\index{HolonomicFunctions package}\index{software package!HolonomicFunctions}%
~\cite{Koutschan10b} 
for \Index{Mathematica}\index{computer algebra system!Mathematica}.

\section{Holonomic and $\d$-Finite Functions}
\label{sec.dfinite}

\index{D-finite|(}\index{$\d$-finite|(}%
In order to state, in an algebraic language, the concepts that are introduced
in this section, and for writing mixed difference-differential equations in a
concise way, the following operator notation is employed: let $\D{x}$ denote
the partial derivative operator with respect to~$x$ ($x$ is then called a
\emph{\Index{continuous variable}}) and $\S{n}$ the forward \Index{shift operator}
with respect to~$n$
($n$ is then called a \emph{\Index{discrete variable}}); they act on a function~$f$ by
\[
  \D{x}f = \frac{\d f}{\d x}
  \quad\text{and}\quad
  \S{n}f = f\big|_{n\to n+1}.
\]
They allow us to write linear homogeneous difference-differential equations
in terms of operators, e.g.,%
\index{linear difference equation}\index{linear recurrence equation}
\[
  \frac{\d}{\d x}f(k,n+1,x,y) + n\frac{\d}{\d y}f(k,n,x,y) + xf(k+1,n,x,y) - f(k,n,x,y) = 0
\]
turns into
\[
  \big(\D{x}\S{n} + n\D{y} + x\S{k} - 1\big) f(k,n,x,y) = 0,
\]
in other words, such equations are represented by polynomials in the operator
symbols $\D{x}$, $\S{n}$, etc., with coefficients in some field~$\F$ which we
assume to be of characteristic~$0$. Note that the polynomial ring
$\F\langle\D{x},\S{n},\dots\rangle$ is not necessarily commutative, a fact
that is indicated by the angle brackets. Its multiplication is subject to the
rules\index{noncommutative multiplication}
\[
  \D{x}\cdot a(x) = a(x)\cdot\D{x} + a'(x)
  \quad\text{and}\quad
  \S{n}\cdot a(n) = a(n+1)\cdot\S{n}.
\]
Typically, $\F$ is a rational function field in the variables $x$, $n$,
etc. over~$\Q$ or over some other field~$\K$.  Such non-commutative rings of
operators\index{operator algebra} were introduced in~\cite{Ore33} and are
called \emph{\Index{Ore algebra}s}.  We use the symbol~$\d$ to denote an
arbitrary operator symbol from an Ore algebra, so that $\d_w$ may stand for
$\S{w}$ or $\D{w}$, for example. Thus, a generic Ore algebra can be written as
$\OO=\F\langle\dd_{\!\w}\rangle$ with, e.g., $\F=\Q(\w)$, where
$\w=w_1,\dots,w_\ell$ and $\dd_{\!\w}=\d_{w_1},\dots,\d_{w_\ell}$.  We define
the \emph{\Index{annihilator}} (w.r.t. some Ore algebra~$\OO$) of a
function~$f$:
\[
  \Ann_{\OO}(f) := \{P\in\OO\mid P(f)=0\}.
\]
It can easily be seen that $\Ann_{\OO}(f)$ is a left \Index{ideal}
in~$\OO$. Every left ideal~$I\subseteq\Ann_{\OO}(f)$ is called an
\emph{\Index{annihilating ideal}} for~$f$.  In the holonomic systems
approach\index{Zeilberger's holonomic systems approach}, functions are
represented by annihilating ideals (plus \Index{initial values}) as a data
structure. When working with left ideals, we use \emph{left \Index{Gr\"obner
    bases}} \cite{Buchberger65, KandrirodyWeispfenning90} which are an
important tool for executing certain operations algorithmically (e.g., for
deciding the ideal membership problem).

\begin{definition}
  Let $\OO=\F\langle\dd_{\!\w}\rangle$ be an \Index{Ore algebra}.  A function~$f$ is
  called \emph{$\d$-finite} or \Index{D-finite} w.r.t.~$\OO$ if
  $\OO/\Ann_{\OO}(f)$ is a finite-dimensional $\F$-vector space. Its dimension
  is called the \emph{\Index{rank}} of~$f$ w.r.t.~$\OO$.
\end{definition}

\begin{example}\label{ex.Laguerre}
Consider the family of \Index{Laguerre polynomials}%
\index{orthogonal polynomials!Laguerre} $L_n^a(x)$ as an example of a
$\d$-finite function w.r.t. $\OO=\Q(n,a,x)\langle\S{n},\S{a},\D{x}\rangle$.
The left \Index{ideal} $I=\Ann_{\OO}(L_n^a(x))$ is generated by the following three
operators that can be easily obtained with the HolonomicFunctions package:%
\index{HolonomicFunctions package}\index{software package!HolonomicFunctions}
\begin{mma}
  \In |Annihilator|[|LaguerreL|[n,a,x],\{|S|[n],|S|[a],|Der|[x]\}]\\
  \Out \{S_{\!a}+D_{\!x}-1,(n+1) S_{\!n}-x D_{\!x}+(-a-n+x-1),x D_{\!x}^2+(a-x+1) D_{\!x}+n\}\\
\end{mma}
\vskip -0.5em\noindent These operators represent well-known identities for
Laguerre polynomials.  Moreover, they are a left Gr\"obner
basis\index{Gr\"obner bases} of~$I$ with respect to the
degree-lexico\-gra\-phic order. Thus, from the leading monomials $\S{a}$,
$\S{n}$, and $\D{x}^2$, one can easily read off that the dimension of the
$\Q(n,a,x)$-vector space $\OO/I$ is two, in other words: $L_n^a(x)$ is
$\d$-finite w.r.t. $\OO$ of \Index{rank}~$2$.
\end{example}

Without proof we state the following theorem about \emph{\Index{closure properties}}
of $\d$-finite functions; its proof can be found in
\cite[Chap. 2.3]{Koutschan09}.  We remark that all of them are algorithmically
executable, and the algorithms work with the above mentioned data structure.

\begin{theorem}\label{thm.clprop}
Let $\OO$ be an \Index{Ore algebra} and let~$f$ and~$g$ be $\d$-finite w.r.t.~$\OO$
of \Index{rank} $r$ and $s$, respectively. Then\nopagebreak[4]\\[-0.5em]
\rule{0pt}{0pt}\hfill\parbox{.98\textwidth}{
\renewcommand{\labelenumi}{(\alph{enumi})}
\begin{enumerate}
\itemsep 0.2em
\item $f+g$ is $\d$-finite of rank $\leq r+s$.
\item $f\cdot g$ is $\d$-finite of rank $\leq r\cdot s$.
\item $Pf$ is $\d$-finite of rank $\leq r$ for any $P\in\OO$.
\item $f|_{x\to A(x,y,\dots)}$ is $\d$-finite of rank $\leq r\cdot d$ if $x,y,\dots$
  are \Index{continuous variable}s and if $A$~satisfies a polynomial equation of
  degree~$d$\index{algebraic functions}.
\item $f|_{n\to A(m,n,\dots)}$ is $\d$-finite of rank $\leq r$ if $A$ is an
  integer-linear expression in the \Index{discrete variable}s $m,n,\dots$.
\end{enumerate}}
\end{theorem}

If we want to consider integration and summation problems, then the function
in question needs to be \emph{holonomic}, a concept that is closely related to
$\d$-finiteness. The precise definition is a bit technical and therefore
skipped here; the interested reader can find it, e.g.,
in~\cite{Zeilberger90,Coutinho95,Koutschan09}. The closure properties for
$\d$-finite functions are also valid for \Index{holonomic functions}.  Additionally,
the following theorem establishes the closure of holonomic functions with
respect to sums and integrals; for its proof, we once again refer
to~\cite{Zeilberger90,Koutschan09}.
\begin{theorem}\label{thm.holo}
  Let the function $f$ be holonomic w.r.t. $\D{x}$ (resp. $\S{n}$). Then also
  $\int_a^b f\rmd x$ (resp. $\sum_{n=a}^b f$) is holonomic.
\end{theorem}
All holonomic functions that appear in this article are also $\d$-finite and
vice versa; therefore we will not continue to care about this subtle
distinction, but only talk about \Index{holonomic functions} from now on. A more
elaborate introduction to holonomic and $\d$-finite functions is given in%
~\cite{Kauers13}.
\index{D-finite|)}\index{$\d$-finite|)}%

\section{Creative Telescoping for Holonomic Functions}
\label{sec.theory}

In order to treat a sum of the form $F(\w)=\sum_{k=a}^bf(k,\w)$%
\index{symbolic summation} with creative telescoping, one has to find an operator~$P$ which
annihilates~$f$, i.e., $Pf=0$, and which is of the form
\begin{equation}\label{eq.ctSum} 
  P=T(\w,\dd_{\!\w})+(\S{k}-1)\cdot C(k,\w,\S{k},\dd_{\!\w})
\end{equation}
where $\dd_{\!\w}$ stands for some operators that act on the variables
$\w=w_1,\dots,w_\ell$. The operator~$T$ is called the \emph{\Index{telescoper}}, and
we will refer to~$C$ as the \emph{\Index{certificate}} or \emph{delta part}. Written
as an equation, \eqref{eq.ctSum} turns into $-Tf(k,\w)=g(k+1,\w)-g(k,\w)$ with
$g(k,\w)=Cf(k,\w)$, compare also with~\eqref{eq.ct1}. With such an
operator~$P$ we can immediately derive a relation for~$F(\w)$:
\begin{align}
0 & = \sum_{k=a}^b P(k,\w,\S{k},\dd_{\!\w}) f(k,\w) \notag \\
  & = \sum_{k=a}^b T(\w,\dd_{\!\w}) f(k,\w)\> +\>\sum_{k=a}^b\big((\S{k}-1)C(k,\w,\S{k},\dd_{\!\w})\big) f(k,\w) \notag \\
  & = T(\w,\dd_{\!\w}) \underbrace{\sum_{k=a}^bf(k,\w)}_{F(\w)}\> +\>
      \underbrace{\vphantom{\sum_{k=a}^b}\Big[C(k,\w,\S{k},\dd_{\!\w}) f(k,\w)\Big]_{k=a}^{k=b+1}}_{\text{inhomogeneous part}}.
  \label{eq.teleSum}
\end{align}
If the \emph{\Index{inhomogeneous part}} evaluates to zero then $T$ is an annihilating
operator for the sum, otherwise we get an inhomogeneous relation. In the
latter case, one can homogenize it by multiplying an annihilating operator for
the inhomogeneous part to~$T$ from the left. Note that in general, the
summation bounds $a$ and $b$ may depend on $\w$ in which case some correction
terms need to be added which are created when the operator~$T$ is pulled in
front of the sum.

In terms of closure properties for \Index{holonomic functions}, see
Theorem~\ref{thm.holo}, this reads as follows: the summand $f(k,\w)$ is given
by an \Index{annihilating ideal} and the operator~$P$ must be a member of this
ideal.  The goal is to compute an annihilating ideal for the function~$F(\w)$
that is sufficiently large (to testify its holonomicity). We have seen that
every operator~$P$ with the above properties yields an annihilating operator
for~$F$, so one continues to compute such creative telescoping operators until
the left \Index{ideal} generated by them is large enough.

Multiple sums\index{summation!multiple sums} can be done by iteratively applying the above procedure.
Alternatively, one can use creative telescoping operators of the form
\begin{equation}\label{eq.ctSumMult}
  T(\w,\dd_{\!\w})+(\S{k_1}-1)\cdot C_1(\k,\w,\SS{\k},\dd_{\!\w})+\dots+(\S{k_j}-1)\cdot C_j(\k,\w,\SS{\k},\dd_{\!\w})
\end{equation}
where $\k=k_1,\dots,k_j$ are the summation variables.

Similarly one derives annihilating operators for an integral\index{symbolic integration}
$I(\w)=\int_a^bf(x,\w)\rmd x$. In this case we look for creative telescoping
operators that annihilate~$f$ and that are of the form
\begin{equation}\label{eq.ctInt}
  P=T(\w,\dd_{\!\w})+\D{x}\cdot C(x,\w,\D{x},\dd_{\!\w}).
\end{equation}
Again, it is straightforward to deduce a relation for the integral
\begin{align}
0 & = \int_a^b P(x,\w,\D{x},\dd_{\!\w}) f(x,\w)\rmd x \notag \\
  & = \int_a^b T(\w,\dd_{\!\w}) f(x,\w)\rmd x \> + \int_a^b \big(\D{x}C(x,\w,\D{x},\dd_{\!\w})\big) f(x,\w)\rmd x \notag \\
  & = T(\w,\dd_{\!\w})\underbrace{\int_a^b f(x,\w)\rmd x}_{I(\w)} \> + \>
      \underbrace{\vphantom{\int_a^b}\Big[C(x,\w,\D{x},\dd_{\!\w}) f(x,\w)\Big]_{x=a}^{x=b}}_{\text{inhomogeneous part}}
  \label{eq.teleInt}
\end{align}
which may be homogeneous or inhomogeneous, as before.  Analogously to the
summation case, multiple integrals can be treated iteratively or by creative
telescoping operators of the form
\begin{equation}\label{eq.ctIntMult}
  T(\w,\dd_{\!\w})+\D{x_1}\cdot C_1(\x,\w,\DD{\x},\dd_{\!\w})+\dots+\D{x_j}\cdot C_j(\x,\w,\DD{\x},\dd_{\!\w}).
\end{equation}
where now $\x=x_1,\dots,x_j$ are the integration variables.

In practice it happens very often that the \Index{inhomogeneous part}
vanishes. The reason for that is because many sums and integrals run over %
\emph{\Index{natural boundaries}}.  This concept is often used, e.g., in
\Index{Takayama's algorithm}, to argue a priori that there will be no
inhomogeneous parts after telescoping. For that purpose, we define that
$\sum_{k=a}^bf$ resp. $\int_a^b f\rmd x$ has natural boundaries if for any
arbitrary operator~$P\in\OO$ for a suitable \Index{Ore algebra}~$\OO$ the
expression $\big[P f\big]_{k=a}^{k=b+1}$ resp.  $\big[P f\big]_{x=a}^{x=b}$
evaluates to zero. Typical examples for natural boundaries are sums with
finite support, or integrals over the whole real line that involve something
like $\exp(-x^2)$.  Likewise contour integrals along a closed path do have
natural boundaries.

\section{Algorithms for Computing Creative Telescoping Relations}
\label{sec.algo}

In this section some algorithms for computing creative telescoping relations
are described briefly; for a detailed exposition see~\cite{Koutschan09}. We
focus on algorithms that are applicable to general \Index{holonomic functions}
and omit those which are designed for special cases of holonomic
functions---like rational, \Index{hypergeometric}, or \Index{hyperexponential}
functions---and refer to Section~\ref{sec.history} and the references given
there. In the following, the summation and integration variables are denoted
by $\bm{v}=v_1,\dots,v_j$ whereas $\w=w_1,\dots,w_\ell$ are the surviving
parameters. So the most general case to consider is a holonomic function
$f(\bm{v},\w)$ which has to be summed and integrated several times, thus some
of the $\bm{v}$ may be discrete variables and the others continuous ones. The
task is to find operators in the (given) annihilating ideal of~$f$ which can
be written in the form
\begin{equation}\label{eq.ctGen}
  T(\w,\dd_{\!\w}) + \Delta_{v_1}\cdot C_1(\bm{v},\w,\dd_{\!\bm{v}},\dd_{\!\w}) + \dots + \Delta_{v_j}\cdot C_j(\bm{v},\w,\dd_{\!\bm{v}},\dd_{\!\w})
\end{equation}
where $\Delta_v=\S{v}-1$ if $v$ is a \Index{discrete variable} and
$\Delta_v=\D{v}$ if $v$ is a \Index{continuous variable}; compare
also with~\eqref{eq.ctSumMult} and~\eqref{eq.ctIntMult}.

\subsection{Zeilberger's Slow Algorithm}
\label{sec.slow}

\index{Zeilberger's slow algorithm}%
In~\cite{Zeilberger90} Zeilberger suggested to approach holonomic sums or
integrals by finding operators whose coefficients are completely free of the
summation and integration variables~$\bm{v}$.  Once such an operator is found,
it is immediate to rewrite it into the form~\eqref{eq.ctGen} using division
with remainder, since the corresponding operators $\dd_{\!\bm{v}}$ now commute
with all remaining variables~$\w$ and with all other
operators~$\dd_{\!\w}$. The theory of holonomic $D$-modules answers the
question whether this elimination is possible at all in an affirmative
way. The same argument justifies the termination of all other algorithms
described in this section. Operators that are free of some variables can be
found, e.g., by a Gr\"obner basis\index{Gr\"obner bases} computation in
$\K(\w)[\bm{v}]\langle\dd_{\!\bm{v}},\dd_{\!\w}\rangle$ or by ansatz and
coefficient comparison. In any case, this algorithm searches for creative
telescoping operators that are not as general as possible---also the
\Index{certificate}s are free of~$\bm{v}$ in contrast to what is indicated
in~\eqref{eq.ctGen}---and therefore is very slow in practice and often does
not find the minimal \Index{telescoper}.

\subsection{Takayama's Algorithm}
\label{sec.takayama}

In order to avoid the overhead that results in a complete elimination of the
$\bm{v}$, Takayama came up with an algorithm that he termed an ``infinite
dimensional analog of Gr\"obner basis''~\cite{Takayama90b}.  He formulated it
only in the differential setting and in a quite theoretical fashion. Chyzak
and Salvy~\cite{ChyzakSalvy98} later presented optimizations that are relevant
in practice and extended it to the more general setting of Ore operators.
Compared to Zeilberger's slow algorithm, \Index{Takayama's algorithm} is
faster and delivers better results, i.e., larger annihilating ideals.

The idea in a nutshell is the following: while in %
\Index{Zeilberger's slow algorithm} first the $\bm{v}$ were eliminated and
then the certificates were divided out, the order is now reversed.  In
Takayama's algorithm one first reduces modulo the right
ideals~$\d_{v_1}\OO,\dots,\d_{v_j}\OO$ and then performs the elimination of
the~$\bm{v}$. The consequence is that the certificates $C_1,\dots,C_j$ are not
computed at all because everything that would contribute to them is thrown
away in the first step. Hence one has to assume a priori that the
\Index{inhomogeneous part}s vanish, e.g., in the case of %
\Index{natural boundaries}. 

There is one technical complication in this approach: one starts with a left
ideal and then divides out some right \Index{ideal}s. After that there is no
ideal structure any more and therefore, one is not allowed to multiply by
either of the variables~$\bm{v}$ from the left.  In order to solve this
problem one enlarges, at the very beginning, the set of generators of the
input annihilating ideal by some of their left multiples by $\bm{v}$-powers
and, at the end, computes a Gr\"obner basis\index{Gr\"obner bases} w.r.t. to
POT ordering (position over term) in the module that is generated by the power
products of~$\bm{v}$.

\subsection{Chyzak's Algorithm}
\label{sec.chyzak}

Chyzak presented his algorithm~\cite{Chyzak00} as an extension of
\Index{Zeilberger's algorithm} to general \Index{holonomic functions}.
Like the latter, \Index{Chyzak's algorithm} can only find creative telescoping
operators for single sums or single integrals. Hence the goal is to find
operators of the form
\begin{equation}
  T(\w,\dd_{\!\w}) + \Delta_v\cdot C(v,\w,\d_v,\dd_{\!\w})
\end{equation}
in the annihilating
ideal~$I\subseteq\K(v,\w)\langle\d_{\!v},\dd_{\!\w}\rangle$ of the
summand or integrand~$f(v,\w)$.  The idea of the algorithm is to make an
ansatz with undetermined coefficients for $T$ and~$C$. Since we may assume
that $C$ is in normal form w.r.t.~$I$, its ansatz is as follows:
\begin{equation}\label{eq.ansatz}
  C(v,\w,\d_v,\dd_{\!\w}) = c_1(v,\w)U_1 + \dots + c_r(v,\w)U_r
\end{equation}
where $U_1,\dots,U_r$ are the monomials which cannot be reduced by~$I$.
Given a Gr\"obner basis\index{Gr\"obner bases} for~$I$, these are exactly the
monomials under its staircase and $r$ is the rank of~$f$. The ansatz for~$T$
is of the form
\begin{equation}\label{eq.teleChy}
  T(\w,\dd_{\!\w}) =  t_1(\w)\dd_{\!\w}^{\aalpha_1} + \dots + t_s(\w)\dd_{\!\w}^{\aalpha_s}
\end{equation}
where $\aalpha_i\in\N^\ell$ for $1\leq i\leq s$. The ansatz $T+\Delta_v\cdot
C$ is reduced with the Gr\"obner basis of~$I$ which leads to a system of
equations for the unknown rational functions $c_1,\dots,c_r,t_1,\dots,t_s$. In
the summation (resp. integration) case, this is a parametrized linear
first-order system of difference (resp. differential) equations in the unknown
functions $c_1\dots,c_r$ and with parameters $t_1,\dots,t_s$. One has to find
rational function solutions of this system and for the parameters, a problem
for which several algorithms exist.  Finally, \Index{Chyzak's algorithm}
proceeds by increasing the support of~$T$ in~\eqref{eq.teleChy} until the
ansatz yields a solution; doing this in a certain systematic way guarantees
that the computed telescopers form a Gr\"obner basis in
$\K(\w)\langle\dd_{\!\w}\rangle$.

\subsection{A Heuristic Approach}
\label{sec.heur}

In~\cite{Koutschan10c} a variant of \Index{Chyzak's algorithm} was developed
that is based on a refined ansatz for the unknown rational functions
$c_1,\dots,c_r$.  The motivation comes from the fact that the bottleneck in
Chyzak's algorithm is to solve the coupled first-order system.  The key
observation is that good candidates for the denominators of the $c_i$ can be
obtained from the leading coefficients of the input Gr\"obner basis.  Thus the
ansatz~\eqref{eq.ansatz} is refined in the following way:
\[
  c_i(v,\w) = \frac{c_{i,0}(\w) + c_{i,1}(\w)v + \dots + c_{i,e_i}(\w)v^{e_i}}{d_i(v,\w)}, \quad 1\leq i\leq r,
\]
where the $d_i$ are explicit polynomials and the $e_i$ are degree bounds for
the numerator; both quantities are determined heuristically. In many examples
this approach is faster than Chyzak's algorithm, but due to its heuristics it
may not always succeed. Note also that this approach can be generalized to
multiple sums and integrals, see Section~\ref{sec.doublesum}.

\section{Demonstration of the HolonomicFunctions Package}
\label{sec.demo}

\subsection{Differential Equations for Bivariate Hypergeometric Functions}
\label{sec.hypergeo2}

\index{HolonomicFunctions package|(}\index{software package!HolonomicFunctions|(}%
The most studied concept in the area of \Index{special functions} are hypergeometric
functions, whose most prominent representative is the Gauss hypergeometric
function~${}_2F_1$. We consider here the Appell hypergeometric function~$F_1$%
\index{Appell functions}\index{generalized hypergeometric functions}%
\index{hypergeometric!generalized functions}
defined by\index{Pochhammer symbol}
\begin{equation}\label{eq.appell1}
  F_1(\alpha,\beta,\beta',\gamma;x,y)=\sum_{m=0}^\infty \sum_{n=0}^\infty
   \frac{x^m y^n (\beta)_m (\beta')_n (\alpha)_{m+n}}{m! n! (\gamma)_{m+n}}
\end{equation}
for $|x|<1$ and $|y|<1$.  Classical mathematical tables like
\cite{GradshteynRyzhik07} list systems of \Index{differential equation}s for such
functions, e.g., entry 9.181 for the Appell functions.  The nature of this
example is that no closed form is desired, but a system of partial
differential equations.  These equations are now derived completely
automatically from~\eqref{eq.appell1} using \Index{Takayama's algorithm}.

The input for Takayama's algorithm is an \Index{annihilating ideal} for the
summand which is obtained by the command \cmd{Annihilator}. We need to
introduce the \Index{shift operator}s $\S{m}$ and $\S{n}$ for the summation
variables and the partial derivatives $\D{x}$ and $\D{y}$ since we are
interested in PDEs w.r.t. $x$ and~$y$. The computation of the annihilating
ideal is direct since the summand is \Index{hypergeometric} in all \Index{discrete variable}s and
\Index{hyperexponential} in all \Index{continuous variable}s:
\begin{mma}
  \In |ann|=|Annihilator|\big[|Pochhammer|[\alpha ,m+n]\,*\,|Pochhammer|[\beta ,m]\,*\linebreak
      |Pochhammer|[b,n]\,/\,(|Pochhammer|[\gamma ,m+n]\,*\,m!\,*\,n!)\,*\,x\pow m\,*\,y\pow n,\linebreak
      \{|S|[m],\,|S|[n],\,|Der|[x],\,|Der|[y]\}\big]\\
  \Out \big\{y D_{\!y}-n,x D_{\!x}-m,\linebreak
    \phantom{\big\{}(mn+m+n^2+n\gamma+n+\gamma) S_{\!n}-(bmy+bny+by\alpha+mny+n^2y+ny\alpha),\linebreak
    \phantom{\big\{}(m^2+mn+m\gamma+m+n+\gamma) S_{\!m}-(m^2x+mnx+mx\alpha+mx\beta+nx\beta+x\alpha\beta)\big\}\\
\end{mma}
\vskip -0.5em Next the double summation is performed and a Gr\"obner
basis\index{Gr\"obner bases} for the left \Index{ideal} containing partial
\Index{differential equation}s satisfied by the series $F_{1}$ is computed:
\begin{mma}
  \In |pde|=|Takayama|[|ann|,\{m,n\}]\\
  \Out \big\{(xy^2-xy-y^3+y^2) D_{\!y}^2+(bx^2-bx) D_{\!x}+(bxy-by^2+xy\alpha-xy\beta+{}\linebreak 
    \kern+25pt xy+x\beta-x\gamma-y^2\alpha-y^2+y\gamma) D_{\!y}+(bx\alpha-by\alpha),\linebreak
    \phantom{\big\{}(x-y) D_{\!x} D_{\!y}-b D_{\!x}+\beta D_{\!y},\phantom{x^2}\linebreak
    \phantom{\big\{}(x^3-x^2y-x^2+xy) D_{\!x}^2+(bxy-by+x^2\alpha+x^2\beta+x^2-xy\alpha-xy\beta-{}\linebreak
    \kern+25pt xy-x\gamma+y\gamma) D_{\!x}+(y\beta-y^2\beta) D_{\!y}+(x\alpha\beta-y\alpha\beta)\big\}\\
\end{mma}
\vskip -0.5em\noindent
Observe that the two equations given in \cite[9.181]{GradshteynRyzhik07} do
not appear in the above result.  To verify that they are nevertheless correct,
one has to show that they are members of the derived \Index{annihilating
  ideal}. This is achieved by reducing them with the Gr\"obner basis\index{Gr\"obner bases} and check
whether the remainder is zero:
\begin{mma}
  \In |OreReduce|[(x(y-1)) |**| (|Der|[x] |Der|[y]) + (y(y-1)) |**| |Der|[y]^2 + \linebreak
      (bx) |**| |Der|[x] + (y(\alpha+b+1)-\gamma) |**| |Der|[y]+\alpha b, \> |pde|]\\
  \Out 0\\
\end{mma}
\vskip -0.5em

On the other hand, the desired equations can be produced automatically by
observing that the first is free of~$\beta'$ and the second does not
involve~$\beta$.  The command \cmd{FindRelation} finds operators in a given
annihilating \Index{ideal} that satisfy certain properties, to be specified by
options:
\begin{mma}
  \In |FindRelation|[|pde|,\,|Eliminate|\to \beta ]\\
  \Out \{(xy-x) D_{\!x} D_{\!y}+(y^2-y) D_{\!y}^2+bx D_{\!x}+(by+y\alpha+y-\gamma) D_{\!y}+b\alpha\}\\
\end{mma}
\vskip -0.5em\noindent This is precisely the form in which the first partial
\Index{differential equation} appears in~\cite{GradshteynRyzhik07} and an
analogous computation yields the second one.

\subsection{An Integral Involving Chebyshev Polynomials}
\label{sec.chebyshev}

It has been pointed out that creative telescoping does not deliver closed-form
solutions. The next example demonstrates how it can be used to prove an
identity, in this case the evaluation of a definite integral%
\index{symbolic integration} which appears in~\cite[7.349]{GradshteynRyzhik07}:
\begin{equation}\label{eq.Tint}
  \int_{-1}^1 \big(1-x^2\big)^{-1/2}\,T_n\big(1-x^2 y\big) \rmd x = \frac{\pi}{2} \big(P_{n-1}(1-y)+P_n(1-y)\big).
\end{equation}
Here $T_{n}(x)$ denotes the \Index{Chebyshev polynomials} of the first kind
defined by
\[
  T_{n}(x) = \cos(n\arccos x)
\]
and the evaluation is given in terms of \Index{Legendre polynomials}%
\index{orthogonal polynomials!Legendre}~$P_{n}(x)$
defined by\index{Rodrigues formula}
\[
  P_{n}(x) = \frac{1}{2^nn!}\frac{\mathrm{d}^n}{\mathrm{d}x^n}\left(x^2-1\right)^n.
\]
This relatively simple example is chosen not only to demonstrate Chyzak's
algorithm but also to enlighten the concept of \Index{closure properties}.

The starting point is the computation of an \Index{annihilating ideal}\index{ideal} for the
integrand $f(n,x,y)=(1-x^2)^{-1/2}\,T_n(1-x^2 y)$ in~\eqref{eq.Tint} which, in
this instance, we will discuss in some more detail. For this purpose, recall
the three-term \Index{recurrence}\index{orthogonal polynomials!three-term recurrence relation}
\begin{equation}\label{eq.Trec}
  T_{n+2}(z) - 2zT_{n+1}(z) + T_n(z) = 0
\end{equation}
and the second-order \Index{differential equation}
\begin{equation}\label{eq.Tdiff}
  (z^2-1) T_n''(z) + zT_n'(z) - n^2T_n(z) = 0
\end{equation}
\index{HolonomicFunctions package}\index{software package!HolonomicFunctions}%
for the Chebyshev polynomials which are both classic and well-known. The
HolonomicFunctions package has these relations stored in a kind of
database. Clearly, the integrand~$f$ also satisfies the
recurrence~\eqref{eq.Trec} if $z$ is replaced by $1-x^{2}y$.  The same
substitution is performed in~\eqref{eq.Tdiff} and considering~$T_n(1-x^2y)$ as
a function in~$y$ yields
\[
  \frac{(1-x^2y)^2-1}{x^4}\frac{\d^2}{\d y^2} T_n(1-x^2y) + \frac{1-x^2y}{-x^2}\frac{\d}{\d y}T_n(1-x^2y) - n^2T_n(1-x^2y) = 0.
\]
Multiplying with $x^2$ produces another annihilating operator
\[
  (x^2y^2-2y)D_{\!y}^2+(x^2y-1)D_{\!y}-n^2x^2
\]
for the integrand~$f$. Note that the square root term can be ignored since it
is free of~$y$. Finally, observe that
\begin{eqnarray*}
  \frac{df}{dx} & = & \frac{-2xy}{\sqrt{1-x^2}}T'_n(1-x^2y) + 
\frac{x}{(1-x^2)^{3/2}}T_n(1-x^2y)\\
  \frac{df}{dy} & = & \frac{-x^2}{\sqrt{1-x^2}}T'_n(1-x^2y)
\end{eqnarray*}
giving rise to the operator
\[
  xD_{\!x}-2yD_{\!y}-\frac{x^2}{1-x^2}
\]
which also annihilates~$f$. The above ad hoc derivation of annihilating
operators for a compound expression can be turned into an algorithmic
method, and this is implemented in the \cmd{Annihilator} command:
\begin{mma}
  \In |Annihilator|\big[|ChebyshevT|[n,1-x^2y]/|Sqrt|[1-x^2],\,\{|S|[n],|Der|[x],|Der|[y]\}\big]\\
  \Out \big\{(x^3-x) D_{\!x}+(2y-2x^2y) D_{\!y}+x^2,\linebreak
       \phantom{\big\{}n S_{\!n}+(x^2y^2-2y) D_{\!y}+(nx^2y-n),\linebreak
       \phantom{\big\{}(x^2y^2-2y) D_{\!y}^2+(x^2y-1) D_{\!y}-n^2x^2\big\}\\
\end{mma}
\vskip -0.5em

\noindent
The above operators form a left Gr\"obner basis\index{Gr\"obner bases}, and therefore differ slightly
from the ones that were derived by hand; but the latter can be obtained as simple
linear combinations of the previous ones.

Now we are ready to perform creative telescoping: we apply \Index{Chyzak's algorithm}
to find operators of the form $T_i+\D{x}C_i$ in the \Index{annihilating ideal}\index{ideal}. Our
implementation returns two such operators, with the property that
$\{T_1,T_2\}$ is a Gr\"obner basis\index{Gr\"obner bases}:
\begin{mma}
  \In \{\{T_1,T_2\},\{C_1,C_2\}\} = |CreativeTelescoping|[\%,\,|Der|[x]]\\
  \Out \bigg\{\big\{(2n^2+2n) S_{\!n}+(2ny^2-4ny+y^2-2y) D_{\!y}+(2n^2y-2n^2+ny-2n),\linebreak
       \phantom{\bigg\{\big\{}(y^2-2y) D_{\!y}^2+(y-2) D_{\!y}-n^2\big\},\linebreak
       \phantom{\bigg\{}\bigg\{\frac{y\left(x^4y-x^2y-2x^2+2\right)}{x} D_{\!y}+y\left(nx^3-nx\right),\frac{x^2-1}{x} D_{\!y}
         \bigg\}\bigg\}\\
\end{mma}

\noindent
With the help of \Index{Mathematica}\index{computer algebra system!Mathematica}, it is
easily verified that the \Index{inhomogeneous part}, see~\eqref{eq.teleInt}, vanishes:
\begin{mma}
  \In |Limit|\big[|ApplyOreOperator|\big[C_1,|ChebyshevT|[n,1-x^2y]/|Sqrt|[1-x^2]\big],x\to 1\big]\\
  \Out 0\\
\end{mma}
\vskip -0.5em

\noindent
(Similar checks have to be done for the lower bound and for $C_2$.)  It
follows that~$T_1$ and~$T_2$ generate an \Index{annihilating ideal}\index{ideal} for the
integral. For the convenience of the user, all the previous steps can be
performed at once by typing a single command:
\begin{mma}
  \In |Annihilator|\big[|Integrate|[|ChebyshevT|[n,1-x^2y]/|Sqrt|[1-x^2],\{x,-1,1\}],\linebreak
      \{|S|[n],|Der|[y]\}\big]\\
  \Out \big\{(2n^2+2n) S_{\!n}+(2ny^2-4ny+y^2-2y) D_{\!y}+(2n^2y-2n^2+ny-2n),\linebreak
       \phantom{\big\{}(y^2-2y) D_{\!y}^2+(y-2) D_{\!y}-n^2\big\}\\
\end{mma}
\vskip -0.5em

The next step is to compute an annihilating ideal for the right-hand side
of~\eqref{eq.Tint}. Instead of applying the \cmd{Annihilator} command to the
expression itself which would produce an annihilating ideal of \Index{rank}~$4$ by
assertion (a) of Theorem~\ref{thm.clprop}, the fact that the sum of the two
\Index{Legendre polynomials}\index{orthogonal polynomials!Legendre} can be
written as $Q(P_{n-1}(1-y))$ with $Q=\S{n}+1$ is employed. This observation
produces an \Index{annihilating ideal}\index{ideal} of rank~$2$, see part (c)
of Theorem~\ref{thm.clprop}:
\begin{mma}
  \In |rhs|=|Annihilator|\big[|ApplyOreOperator|[|S|[n]+1,|LegendreP|[n-1,1-y]],\linebreak
      \{|S|[n],|Der|[y]\}\big]\\
  \Out \big\{(2n^2+2n) S_{\!n}+(2ny^2-4ny+y^2-2y) D_{\!y}+(2n^2y-2n^2+ny-2n),\linebreak
       \phantom{\big\{}(y^2-2y) D_{\!y}^2+(y-2) D_{\!y}-n^2\big\}\\
\end{mma}
\vskip -0.5em

Finally, one realizes that the annihilating ideals for both sides of the
identity coincide. The proof is completed by comparing two \Index{initial values},
e.g., for $n=0$ and $n=1$. This has to be done by hand (of course, with the
help of the computer algebra system), but is not part of the functionality of
the HolonomicFunctions package.

\subsection{A q-Holonomic Summation Problem from Knot Theory}

The \Index{colored Jones function} is a powerful knot invariant; it is a \Index{$q$-holonomic}
sequence of Laurent polynomials~\cite{GaroufalidisLe05}.  Its \Index{recurrence}
equation is of interest since it seems to be closely related with the
$A$-polynomial of a knot. The recurrence for the colored Jones function
$J_{7_4,n}(q)$ of the knot $7_4$ was derived in~\cite{GaroufalidisKoutschan13}
using creative telescoping, starting from the sum representation
\begin{equation}\label{eq.CJ74}
  J_{7_4,n}(q) = \sum_{k=0}^{n-1} (-1)^k \big(c_k(q)\big)^2 q^{-k n -\frac{k(k+3)}{2}} (q^{n-1};q^{-1})_k (q^{n+1};q)_k 
\end{equation}
where $(x;q)_n$ denotes the \Index{$q$-Pochhammer symbol} defined as
$\prod_{j=0}^{n-1}(1-xq^j)$ and where the sequence $c_k(q)$ satisfies a
second-order \Index{recurrence}:
\begin{equation}\label{eq.ck}
  c_{k+2}(q) + (q^{k+3}+q^{k+4}-q^{2 k+5}+q^{3 k+7})c_{k+1}(q) + (q^{2 k+6}-q^{3 k+7})c_k(q) = 0.
\end{equation}
Note that the summand in~\eqref{eq.CJ74} is not
\Index{$q$-hypergeometric}\index{summation!$q$-holonomic}\index{summation!holonomic}
and therefore the $q$-version of \Index{Zeilberger's algorithm} cannot be applied.

Again, we start by constructing an \Index{annihilating ideal}\index{ideal} for the summand.  The
one for the sequence $c_k(q)$ is given by its definition~\eqref{eq.ck}, we
just have to add the trivial relation w.r.t.~$n$ and convert everything to
operator form (note the usage of \Index{$q$-shift operator}s):
\begin{mma}
  \In |annc| = |ToOrePolynomial|\big[\big\{|QS|[|qn|, q^n]-1, |QS|[|qk|, q^k]^2 + \linebreak
    \big(q^{k+3}(1 + q - q^{k+2} + q^{2k+4})\big)|**||QS|[|qk|, q^k] + q^{2k+6}(1-q^{k+1})\big\}\big] \\
  \Out \big\{S_{\!\text{qn},q}-1, 
    S_{\!\text{qk},q}^2 + \big(q^7\text{qk}^3-q^5\text{qk}^2+q^4\text{qk}+q^3\text{qk}\big) S_{\!\text{qk},q}+
    \big(q^6\text{qk}^2-q^7\text{qk}^3\big)\big\}\\
\end{mma}
\vskip -0.5em\noindent
Next, the closure property ``multiplication'', see Theorem~\ref{thm.clprop} (b),
is applied (the result is about 2 pages long and therefore not displayed
here):
\begin{mma}
  \In |annSmnd| = |DFiniteTimes|[|annc|, |annc|, \linebreak 
      |Annihilator|[(-1)^k \, q\pow (-kn - k(k + 3)/2) \> |QPochhammer|[q^{n-1}, 1/q, k] \linebreak
      |QPochhammer|[q^{n+1}, q, k], \> \{|QS|[|qk|, q^k], |QS|[|qn|, q^n]\}]];\\
\end{mma}
\bigskip

The stage is now prepared for calling \Index{Chyzak's algorithm} which
delivers a pair $(T,C)$ consisting of \Index{telescoper} and \Index{certificate}:
\begin{mma}
  \In \{T,C\} = |CreativeTelescoping|[|annSmnd|, \> |QS|[|qk|, q^k] - 1]\\
\end{mma}
\medskip\noindent
This computation takes about two minutes and the result is again too large to
be printed here. We remark that the \Index{inhomogeneous part} does not vanish so that
we obtain an inhomogeneous \Index{recurrence} for the function $J_{7_4,n}(q)$.  The
result is in accordance with the AJ conjecture and the previously known
$A$-polynomial of the knot~$7_4$.

\subsection{A Double Integral Related to \Index{Feynman Diagrams}}
\label{sec.feynman}

We study the double integral
\begin{equation}\label{eq.feyn}
  \int_0^1\int_0^1\frac{w^{-1-\varepsilon /2}(1-z)^{\varepsilon /2}z^{-\varepsilon /2}}{(z+w-wz)^{1-\varepsilon}}
  \left(1-w^{n+1}-(1-w)^{n+1}\right)\rmd w\rmd z
\end{equation}
than can be found in~\cite[(J.17)]{Klein06}. The task is to compute a
\Index{recurrence} in~$n$ where~$\varepsilon$ is just a parameter. We are aware of the fact that~\eqref{eq.feyn}
is not a hard challenge for physicists, and we use it only as a proof of
concept here. We are going to apply \Index{Chyzak's algorithm} iteratively.

For computing an \Index{annihilating ideal}\index{ideal} for the inner integral, we simply use the
command \cmd{Annihilator} that takes care of the \Index{inhomogeneous part}
automatically:
\begin{mma}
 \In f=w\pow(-1-\varepsilon/2)\,(1-z)\pow(\varepsilon/2)\,z\pow(-\varepsilon/2)/(w+z-w\,z)\pow(1-\varepsilon)\linebreak
     (1-w\pow(n+1)-(1-w)\pow(n+1));\\
\end{mma}
\smallskip
\begin{mma}
 \In |ann|=|Annihilator|\big[|Integrate|[f,\{w,0,1\}],\:\{|S|[n],\,|Der|[z]\}\big];\\
\end{mma}
\vskip 0.5em\noindent This result is quite large so that we do not want to display it here.
But it can be used again as input to \Index{Chyzak's algorithm}, in order to
treat the outer integral.
\begin{mma}
 \In \{\{T\},\{C\}\}=|CreativeTelescoping|[|ann|,\:|Der|[z],\:|S|[n]];\\
\end{mma}
\medskip\noindent It is a little bit tricky to handle the \Index{inhomogeneous part}
of the outer integral since it involves an integral itself:
\begin{equation}\label{eq.feynInh}
  \left[C \int_0^1 f\rmd w\right]_{z=0}^{z=1}=
  \int_0^1\big[Cf\big]_{z=0}^{z=1}\,\rmd w.
\end{equation}
It turns out that the right-hand side of~\eqref{eq.feynInh} is preferable to
show that the \Index{inhomogeneous part} evaluates to zero.  Therefore the
operator~$T$ annihilates the double integral, and this is the desired
\Index{recurrence} in~$n$ (which is of order~$3$):
\begin{mma}
 \In |Factor|[T]\\
 \Out -(\varepsilon -n-3)(\varepsilon -n-2)(\varepsilon +2n+4)(\varepsilon +2n+6)S_{\!n}^3+\linebreak
      (\varepsilon -n-2)(\varepsilon +2n+4)(\varepsilon ^2+2\varepsilon n+5\varepsilon -6n^2-28n-34)S_{\!n}^2-\linebreak
      (n+2)(\varepsilon^3\!-3\varepsilon^2n-6\varepsilon^2\!-8\varepsilon n^2\!-30\varepsilon n-28\varepsilon +12n^3+64n^2+116n+72)S_{\!n}-\linebreak
      2(n+1)(n+2)^2(\varepsilon -2n-2)\\
\end{mma}

\subsection{A Hypergeometric Double Sum}
\label{sec.doublesum}

We finally turn to a binomial double sum which was investigated
in~\cite{AndrewsPaule93}\index{binomial coefficient}:
\begin{equation}\label{eq.APsum}
  \sum_i\sum_j\binom{i+j}{i}^{\!\!2}\binom{4n-2i-2j}{2n-2i}=(2n+1)\binom{2n}{n}^2.
\end{equation}
We apply the heuristic approach from Section~\ref{sec.heur} to it.  The
corresponding command in the HolonomicFunctions package is
\cmd{FindCreativeTelescoping}:
\begin{mma}
  \In |FindCreativeTelescoping|[|Binomial|[i+j,i]\pow 2 \>\> |Binomial|[4n-2i-2j,2n-2i], \linebreak
      \{|S|[i]-1,|S|[j]-1\}, \> |S|[n]]\\
  \Out \bigg\{\{1\}, \bigg\{\bigg\{\frac{-2 i^2 j+i^2 n-i^2-2 i j^2+3 i j n-2 i j+3 i n}{(j+1) (i+j-2 n)}, \linebreak
       \phantom{\bigg\{\{1\}, \bigg\{\bigg\{} \frac{-2 i^2 j-2 i j^2+3 i j n-2 i j+j^2n-j^2+3 j n}{(i+1) (i+j-2 n)}\bigg\}\bigg\}\bigg\}\\
\end{mma}
The output consists of the telescoper and the two certificates. At first
glance it may seem contradictory that the telescoper is~$1$, but there are
contributions from the certificates that make the recurrence for the double
sum inhomogeneous.  So we don't claim that the operator~$1$ annihilates the
double sum, which would imply that it is zero.
\index{HolonomicFunctions package|)}\index{software package!HolonomicFunctions|)}%

\section{Selected Applications of Creative Telescoping}
\label{sec.appl}

In this section we want to give an extensive, but certainly not complete,
collection of examples which show the beneficial use of creative telescoping
in diverse areas of mathematics and physics.


\Index{Zeilberger's algorithm} for \Index{hypergeometric}
sums\index{summation!hypergeometric} is a meanwhile so classic tool that it is
impossible to list all papers where it has been used to prove some binomial sum
identity\index{binomial coefficient}. We therefore restrict ourselves to
publications where this algorithm plays a more or less central role.  In%
~\cite{EkhadZeilberger94} 
it was used to prove Ramanujan's famous formula for $\pi$, and in%
~\cite{Guillera06} 
for some formulas of similar type.  The whole paper%
~\cite{Strehl94} 
is dedicated to binomial identities that arise in combinatorics and how to
prove them algorithmically.  Two proofs of the notorious binomial double sum
identity~\eqref{eq.APsum} are given in%
~\cite{AndrewsPaule93} 
where, due to the lack of multi-summation software packages at that time, the
problem was reduced in a tricky way to a single sum identity. A ``triumph of
computer algebra'' is celebrated in%
~\cite{Prodinger96} 
where the computation of factorial moments and probability generating
functions for heap ordered trees is based on Zeilberger's algorithm. In%
~\cite{AmdeberhanZeilberger97} 
it is used to derive formulas for hypergeometric series acceleration, among
them a pretty formula for $\zeta(3)$\index{Riemann's zeta function} that
allowed to evaluate this constant to a large number of digits. In the article%
~\cite{Koornwinder98}, 
\Index{Zeilberger's algorithm} is combined with asymptotic estimates in order to give
automated proofs of non-terminating series identities of Saalsch\"utz type.
Applications in the context of orthogonal polynomials%
\index{orthogonal polynomials!and computer algebra} are given in%
~\cite{KoepfSchmersau98}. 
A fast way of computing Catalan's constant is derived in%
~\cite{Zudilin03} 
by means of creative telescoping. While the \Index{recurrence} that plays a
crucial role in Ap\'ery's proof of the irrationality of $\zeta(3)$ is nowadays
a popular example for demonstrating these techniques, they were not available
to Ap\'ery when he came up with his proof\index{Apery sequence}.  A new,
elementary proof, still using \Index{Zeilberger's algorithm}, is given in%
~\cite{Zudilin09}. 
We conclude this paragraph by mentioning%
~\cite{AmdeberhanAngelisLinMollSury12} 
where a binomial identity that arose in the study of a certain integral is
investigated.


We turn to applications of creative telescoping that go beyond Zeilberger's
algorithm.  As an application of its $q$-analogue we cite\index{summation!$q$-hypergeometric}%
~\cite{Paule94} 
where computer proofs for the Rogers-Ramanujan identities are constructed.
Multi-summation techniques\index{summation!multiple sums} for \Index{$q$-hypergeometric} terms were used in%
~\cite{BerkovichRiese02} 
to prove a partition theorem of G\"ollnitz. Computer proofs for summation
identitites involving \Index{Stirling numbers} are given in%
~\cite{KauersSchneider08}. 
In~\cite{BostanChyzakLecerfSalvySchost07} 
creative telescoping was used to obtain bounds on the order and degree of
\Index{differential equation}s satisfied by \Index{algebraic functions}.
\Index{Chyzak's algorithm} was applied to the \Index{generating function} of 3-dimensional
rook paths%
~\cite{BostanChyzakHoeijPech11} 
in order to derive an explicit formula. Creative telescoping proofs for a
selection of special function\index{special functions} identities, mostly involving integrals, are
presented in%
~\cite{KoutschanMoll11}. 
Another application to the evaluation of integrals is%
~\cite{AmdeberhanKoutschanMollRowland12}. 


In~\cite{Zeilberger07} 
Zeilberger proposed an approach how to evaluate determinants of matrices with
holonomic entries with the method of creative telescoping. This approach
applies to determinants of the form $\det_{1\leq i,j\leq n}(a_{i,j})$ whose
entries are bivariate holonomic sequences, not depending on the dimension~$n$.
The so-called ``\Index{holonomic ansatz}'' celebrated its greatest success so
far when it was employed to prove the qTSPP conjecture%
~\cite{KoutschanKauersZeilberger11}, 
a long-standing prominent problem in enumerative combinatorics, which
previously had been reduced to a certain determinant evaluation of the above
type. This conjecture is the $q$-analogue of what is known as Stembridge's
theorem about the enumeration of totally symmetric plane partitions. Based on
creative telescoping, this theorem was re-proved twice, both times using the
formulation as a determinant evaluation: the first time by applying
\Index{symbolic summation} techniques to a decomposition of the matrix%
~\cite{AndrewsPauleSchneider05}, 
the second time following the holonomic ansatz%
~\cite{Koutschan10a}. 
Some extensions of the holonomic ansatz were presented in%
~\cite{KoutschanThanatipanonda13} 
and were applied to solve several conjectures about determinants.  An
analogous method for the evaluation of Pfaffians was developed in%
~\cite{IshikawaKoutschan12}. 


In the field of quantum topology and knot theory, a prominent object of
interest is the so-called \emph{\Index{colored Jones function}} of a knot. This
function is actually an infinite sequence of Laurent polynomials and in%
~\cite{GaroufalidisLe05} 
it has been shown that this sequence is always \Index{$q$-holonomic}, by establishing
an explicit multisum\index{summation!multiple sums} representation with proper \Index{$q$-hypergeometric}
summand\index{summation!$q$-hypergeometric}. The corresponding minimal-order
\Index{recurrence} is called the \emph{non-commutative A-polynomial} of the
knot. Creative telescoping was used to compute it for a family of twist knots%
~\cite{GaroufalidisSun10} 
and for a few double twist knots~\cite{GaroufalidisKoutschan13}.


We are turning to applications in the area of numerical analysis.  A widely
used method for computer simulations of real-world phenomena described by
partial differential equations is the \emph{\Index{finite element method}} (FEM). A
short motivation of using \Index{symbolic summation} techniques in this area is given
in%
~\cite{PaulePillweinSchneiderSchoeberl06}, 
and a concrete application where hypergeometric summation algorithms deliver
certain \Index{recurrence} equations which allow for a fast evaluation of the basis
functions, is described
in~\cite{BecirovicPaulePillweinRieseSchneiderSchoeberl06}.  Further examples,
where creative telescoping is used for verifying identities arising in the
context of FEM or for finding identities that help to speed up the numerical
simulations, can be found in
\cite{BeuchlerPillwein07, 
  BeuchlerPillweinZaglmayr13, 
  KoutschanLehrenfeldSchoeberl12}.


Last but not least we want to point out that creative telescoping has
extensively supported computations in physics. We will not detail on the very
fruitful interaction of summation methods in difference fields with the
computation of Feynman integrals in particle physics%
~\cite{AblingerBluemleinKleinSchneider10}, 
but refer to the survey%
~\cite{Schneider13}, 
and the references therein. The estimation of
the entropy of a certain process%
~\cite{LyonsPauleRiese02} 
was supported by computer algebra. In the study of generalized two-Qubit
Hilbert-Schmidt separability probabilities%
~\cite{Slater13} 
creative telescoping was employed to simplify a complicated expression
involving \Index{generalized hypergeometric functions}%
\index{hypergeometric!generalized functions}. The authors of%
~\cite{BostanBoukraaChristolHassaniMaillard12} 
underline the particular importance that creative telescoping may play in the
evaluation of the $n$-fold integrals $\chi^{(n)}$ of the magnetic
susceptibility of the Ising model.  Also relativistic Coulomb integrals\index{relativistic Coulomb!integrals} have
been treated with the holonomic systems approach\index{Zeilberger's holonomic systems approach}%
~\cite{PauleSuslov13}. 
Likewise it was used in the proof of a third-order integrability criterion for
homogeneous potentials of degree~$-1$~\cite{CombotKoutschan12}.  One branch of
statistical physics deals with random walks on lattices\index{lattice walks}; some results in this
area%
~\cite{ZhangWanLuXu11, 
  Koutschan13} 
were obtained by creative telescoping.%
\index{holonomic|)}%
\index{creative telescoping!holonomic functions|)}


\bibliographystyle{plain}
\bibliography{literatur}


\end{document}